\documentclass[twocolumn,showpacs,unsortedaddress,superscriptaddress,showkeys,nofootinbib,preprintnumbers,letterpaper]{revtex4-1}
\usepackage{acronym}
\usepackage{amsmath}
\usepackage{amssymb}
\usepackage{graphics}
\usepackage{graphicx}
\usepackage{subfigure}
\usepackage[squaren]{SIunits}
\usepackage[usenames]{color}
\usepackage{url}

\newcommand*{\diff}{\,\mathrm{d}}

\newcommand{\Hanford}{\mathrm{H}1}
\newcommand{\Livingston}{\mathrm{L}1}
\newcommand{\likelihood}{\mathcal{L}}

\newcommand{\figref}[1]{Fig.~\ref{#1}}

\addunit{\annum}{a}
\addunit{\AU}{AU}
\addunit{\parsec}{pc}

\begin{document}


\title{Development and Application of a Detection System for a Novel Class of Gravitational-Wave Transients}

\author{Soichiro Kuwahara}
\affiliation{Research Center for the Early Universe (RESCEU), Graduate School of Science, The University of Tokyo, Tokyo 113-0033, Japan}

\author{Kipp Cannon}
\affiliation{Research Center for the Early Universe (RESCEU), Graduate School of Science, The University of Tokyo, Tokyo 113-0033, Japan}
\date{\today}

\begin{abstract}
We previously described the development of a detection system for a novel class of transient gravitational-wave sources taking the form of Cherenkov-like bursts \cite{skuwaharaMA}.
Here, we have applied the system to the data of the LIGO/Virgo/KAGRA O3 science run, and report a null result.
The \textit{ad hoc} waveform model is motivated by the conjectured emission of gravitational waves from a curvature source moving at super-luminal speed, and while there is no plausible natural or artificial source of such waves, we nevertheless use the null result to infer a tongue-in-cheek upper bound on the number density of near-Earth transits of spacecraft travelling at warp speed.
The upper bound is parameterized in terms of the trajectory's impact parameter, the vehicle's engine power, and speed.
These quantities can be connected to statements made in science fiction allowing us to translate the upper bound into a bound on the number density of specific types of spacecraft from, for example, Star Trek or Star Wars.
Although most suitable for entertainment purposes, these constraints might find use being folded into a Bayesian inference type estimate on the number of extra-terrestrial civilizations in the galaxy.
\end{abstract}


\maketitle
\acrodef{GW}{gravitational-wave}
\acrodef{SNR}{signal-to-noise ratio}
\acrodef{PDF}{probability density function}

\section{Introduction}
The Advanced Laser Interferometer Gravitational-Wave Observatory (LIGO) \cite{DetectorALIGO} and Advanced Virgo \cite{DetectorAVirgo} detectors are \ac{GW} detectors.  LIGO and Virgo have completed the third observing run (O3) and they will start the fourth observing run (O4) collaborating with the Kamioka Gravitational-wave detector, Large-scale Cryogenic Gravitational-wave Telescope (KAGRA) \cite{DetectorKAGRA} in 2023.
\ac{GW} transients from the collisions of neutron stars and of black holes are regularly observed by \ac{GW} detectors \cite{GWTC3}.  Searches are also conducted for a variety of other \ac{GW} transients such as bursts from core collapse supernovae \cite{CCSNO1O2,TransientsO3LVK}, cosmic strings \cite{stringsearchnew}, cracking neutron star crusts \cite{TransientsO3LVK}, and so on.
In this study, we have developed a detection system for a novel class of Cherenkov burst-like \ac{GW} transients, and search for evidence of this phenomenon in the data from the LIGO/Virgo/KAGRA O3 science run \cite{GWOSCO3}.

Our prototype source is a spacetime curvature source (\textit{e.g.}, an object with mass) moving at super-luminal speed with respect to surrounding observers.  Previous authors have constructed exact solutions of Einstein's equation in which such behaviour is observed.  One example is the Alcubierre warp drive \cite{Alcubierre1994warp}.  This is a solution in which a spherical shell containing a useful volume of flat spacetime (where a spaceship can be placed) moves faster than light with respect to surrounding observers.

Two notable features of this solution are (i) that it very rapidly becomes flat outside the spherical shell, with no out-going wave components, nor any disturbance whatsoever away from the shell as it passes, and (ii) the construction of the shell requires material with negative rest mass, and no such material is known to exist.  Other solutions of Einstein's equations that exhibit faster-than-light movement of objects share both properties, for example Krasnikov tubes \cite{Krasnikov}, and traversable wormholes \cite{MorrisThorneWormhole}.  We wonder if solutions with an out-going wave component, \textit{i.e.}\ solutions with a wake, might not require negative mass to construct.  Consider a fluid dynamics analogy:  boats move through water faster than the speed of surface waves, and leave wakes behind themselves;  is it possible to design a boat hull with a useful interior volume and that moves through water faster than the speed of surface waves but that leaves no distrubance behind it whatsoever?  One might find that it \emph{is} possible to construct such a solution but only if non-physical materials, for example substances with negative volumes, are used to construct the boat's hull.  Perhaps physically realizable faster-than-light propulsion mechanisms must have out-going wave components, just as real boats produce wakes.

We have considered what form the out-going wave field from such a solution of Einstein's equation might take in the far-field regime.  We do not present a solution of Einstein's equation, and so we have no concrete wave emission mechanism.
We conjecture that the outgoing wake from super-luminal sources will be similar to the waveform of Cherenkov radiation.
Cherenkov radiation is the radiation emitted by a charged particle moving faster than the speed of light in a medium \cite{Cherenkovdiscovery,FrankTamm}.

Previous authors have considered gravitational Cherenkov radiation and Lorentz violation \cite{PreviousSeachSchreck}\cite{PreviousSeachAlan}, however, these studies are deriving constraints on coefficients of Standard Model Extensions by assuming the absence of such phenomena.

We are aware of no previous experimental attempts to directly detect gravitational Cherenkov radiation.

\section{Waveform}
\label{sec:waveform}

\subsection{Cherenkov Radiation}
\label{sec:CherenkovSpectrum}
Cherenkov radiation was experimentally confirmed by Pavel Cherenkov in 1934 \cite{Cherenkovdiscovery}.
It is the bluish glow which occurs when charged particles move faster than the speed of light in a medium.
The energy spectrum of Cherenkov radiation was formulated by Frank and Tamm in 1937 \cite{FrankTamm}.
The radiated energy per unit distance along the path of a particle of charge \(z e\) moving at speed \(\beta\) with respect to light is \cite{Jackson}.
\begin{align}
        \frac{\diff E}{\diff x}=\frac{(ze)^2}{c^2}\int_{\epsilon(\omega)>(1/\beta^2)}\omega\left(1-\frac{1}{\beta^2\epsilon(\omega)}\right)\diff\omega,
\end{align}
where $\epsilon(\omega)$ is the frequency-dependent macroscopic dielectric constant.
Using the relation for refractive index $n(\omega)=\sqrt{\epsilon(\omega)}$, the spectral density becomes
\begin{equation}
\frac{\diff^2E}{\diff x\diff\omega}
   = \begin{cases}
   \frac{(ze)^2}{c^2}\omega\left(1-\frac{1}{\beta^2n(\omega)^2}\right) & \text{for \(\beta n(\omega)>1\)}, \\
   0 & \text{otherwise}.
   \end{cases}
\end{equation}
Radiation occurs in the Cherenkov frequency band, defined by $\beta n(\omega) > 1$.

In this study, since there is no refractive or dispersive medium assumed through which the super-luminal source is travelling, we take $n(\omega) = 1$ to be a constant.  Furthermore, we are not discussing an electromagnetic wave so \(ze\) is meaningless, therefore we cannot derive the magnitude of the spectrum from first principles, we can only discuss it up to an unknown proportionality constant.  In that case, the shape of the spectrum depends on a single parameter, \(\beta\), giving the ratio of the source velocity to the wave speed.
\begin{equation}
\label{eq:FrankTamm}
\frac{\diff^2E}{\diff x\diff\omega}
   \propto \begin{cases}
   \frac{\omega}{c^2}\left(1-\frac{1}{\beta^2}\right) & \text{for \(\beta>1\)}, \\
   0 & \text{otherwise}.
   \end{cases}
\end{equation}

\begin{figure}
\resizebox{\linewidth}{!}{\includegraphics{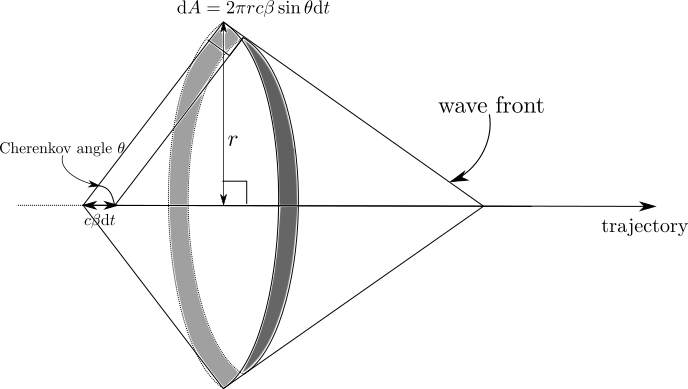}}
\caption{Cherenkov radiation cone picture.}
\label{fig:Cherenkov1r}
\end{figure}

The out-going wake will form a conical wavefront with the source at the vertex.  See \figref{fig:Cherenkov1r}.
The energy emitted by the source during the time interval \(\diff t\) as it travels a distance \(\diff x = c \beta \diff t\) along its trajectory is radiated into the shaded circular segment of the cone's surface, the area of which is
\begin{equation}
\label{eq:ConeRelations}
\diff A
   = 2 \pi r c \beta \sin \theta \diff t
   = 2 \pi r \sin \theta \diff x.
\end{equation}
The Cherenkov angle $\theta$ is determined by the ratio of the speed of the source to the wave speed, and is $\cos\theta = \beta^{-1}$, or \(\sin\theta = \beta^{-1} \sqrt{\beta^{2} - 1}\).

\subsection{Energy of Gravitational Wave}
\label{sec:GWEnergySpectrum}
The spectral flux density of a \ac{GW} is \cite{Maggiore}
\begin{multline}
\label{eq:GWenergy}
\frac{\diff^2E}{\diff A\diff f}
   = \frac{\pi c^3}{2G}f^2\left(|\tilde{h}_{+}(f)|^2+|\tilde{h}_{\times}(f)|^2\right) \\
   = \frac{\pi c^3}{2G}f^2 |\tilde{h}_{+}(f)|^2,
\end{multline}
where, because of the axial symmetry of the problem, we assume the out-going \ac{GW} is linearly polarized, and so have set \(\tilde{h}_{\times}(f) = 0\) (arbitrarily labelling the wave's poarlization as ``\(+\)'').

Combining \eqref{eq:FrankTamm}, \eqref{eq:ConeRelations} and \eqref{eq:GWenergy} gives us
\begin{equation}
\label{eq:unnormalizedspectrum}
|\tilde{h}_{+}(f)|^2
   = B^{2} \frac{G}{c^{4}} \frac{1}{f} \frac{1}{r} \frac{\sqrt{\beta^{2} - 1}}{\beta},
\end{equation}
where we've introduced a proportionality constant \(B^{2}\).

\subsection{Amplitude Normalization and Ultraviolet Divergence}
\label{sec:rho1r}

Using \eqref{eq:GWenergy}, we can relate the wave's amplitude to the energy contained in it, and so we can rewrite the unknown proportionality coefficient in \eqref{eq:unnormalizedspectrum} in terms of the power output of the vehicle's propulsion system.
Here, the power will be left as an additional free parameter of the model, however one should expect that a solution of Einstein's equation for a proper source model would provide an expression for the amplitude of the outgoing wave, and remove this degree of freedom.

If maintaining its speed requires the vehicle's propulsion system to consume a power \(P\), then from \eqref{eq:ConeRelations} the areal flux density of the wave is
\begin{equation}
\label{eq:Efunc}
\frac{\diff E}{\diff A}
   = \frac{P\diff t}{\diff A}
   = \frac{P}{2\pi rc \sqrt{\beta^{2} - 1}}.
\end{equation}

We can equate this to \(\diff E / \diff A\) for the \ac{GW} by substituting \eqref{eq:unnormalizedspectrum} back into \eqref{eq:GWenergy} and integrating over \(f\).  Doing so yields an integrand \(\propto f\), and so the total areal flux density of the \ac{GW} diverges due to high frequency contributions.  We conjecture that a source with non-zero length along the direction of travel will produce a wake with a high-frequency cut-off determined by some combination of the length of the vehicle and \(\beta\).  This will occur due to the destructive superposition of high-frequency contributions to the wake from all points along the length of the vehicle:  each point along the vehicle contributes a wake delayed slightly with respect to the point just ahead of it;  for any given frequency a delay is a phase rotation;  so for each frequency integrating the contributions to the wake along the vehicle becomes an integral over phase; therefore for frequencies corresponding to wavelengths much shorter than the length of the vehicle the net contribution should be approximately 0.

Current \ac{GW} detectors are not sensitive to frequencies higher than a few kilohertz, corresponding to wavelengths of hundreds of kilometres.  We make the assumption that sources of interest are much smaller than this, and therefore for the purposes of this study the details of the high frequency cutoff are irrelevant as they only affect portions of the \ac{GW} spectrum that are inaccessible to the detector.  We choose, therefore, to implement the high-frequency cut-off as a simple hard cut-off at some frequency
\begin{equation}
f_{\mathrm{cutoff}}
   = c / \ell_{\mathrm{cutoff}},
\end{equation}
an additional free parameter of the model.

Combining \eqref{eq:GWenergy} and \eqref{eq:unnormalizedspectrum}, and integrating up to \(f_{\mathrm{cutoff}}\),
\begin{align}
\label{eq:eqn8}
\frac{\diff E}{\diff A}
   & = B^{2} \frac{\pi}{4 c} \frac{1}{r} \frac{\sqrt{\beta^{2} - 1}}{\beta} f_{\mathrm{cutoff}}^{2}.
\end{align}
Equating this to \eqref{eq:Efunc} allows us to obtain an expression for the amplitude normalization in terms of the power output
\begin{equation}
\label{eq:normalization}
B^{2}
   = \frac{2 \beta P}{\pi^{2} f_{\mathrm{cutoff}}^{2} (\beta^{2} - 1)}.
\end{equation}
The final, normalized, magnitude of the waveform in the frequency domain is
\begin{multline}
\label{eq:waveform}
|\tilde{h}_{+}(f)|
   = \sqrt{\frac{4 G}{\pi c^{3} f_{\mathrm{cutoff}}^{2}} \frac{\diff E}{\diff A} \frac{1}{f}}
   = \sqrt{\frac{2 G P \ell_{\mathrm{cutoff}}^{2}}{\pi^{2} c^{6} \sqrt{\beta^{2} - 1}} \frac{1}{f} \frac{1}{r}}
   \\
   = \sqrt{\frac{2 G P}{\pi^{2} c^{4} f_{\mathrm{cutoff}}^{2} \sqrt{\beta^{2} - 1}} \frac{1}{f} \frac{1}{r}}
\end{multline}

Note that the flux density in \eqref{eq:Efunc}, or equivalently \eqref{eq:eqn8}, falls off with distance as \(r^{-1}\), whereas usually radiation in the far-field regime falls off with distance as \(r^{-2}\).  The conical wavefront maintains a higher amplitude for distant observers than one would expect from energy radiated spherically.  This will play a significant role in the construction a detection algorithm for this class of waveform.  This feature of Cherenkov radiation has been confirmed experimentally \cite{CherenkovRealData}.  See, specifically, \cite[Fig.~3]{CherenkovRealData} where $\delta z$ represents a distance from the source to the observer.

\subsection{Causality and Infrared Divergence}
\label{sec:template}

Two problems remain to be addressed before we have an explicit form for the waveform:  we only know the magnitude of the waveform in the frequency domain, we don't know the phases of the frequency components, and while we have removed an energy density divergence at high frequencies there remains a divergence in the strain amplitude at low frequencies.

Arbitrarily low frequency \acp{GW} are non-physical:  they carry arbitrarily little energy and are indistiguishable from a fixed background spacetime.  High amplitude low frequency components make the strain waveform look impressive when plotted but they contribute nothing to the detectability of the signal.  To simplify the numerical operations required to generate and process simulated waveforms, we set all frequency components below a low-frequency cut-off to 0.  We choose this cut-off to be \(\unit{10}{\hertz}\), which is the lowest frequency to which the current generation of \ac{GW} detectors are calibrated.

We obtain the phases for the frequency components by requiring the time domain waveform to be causal.
We do this by first assuming the phases are all \(\unit{0}{\radian}\), and inverse Fourier transforming to the time domain.
At this stage the waveform is shaped like a narrow spike, with a tip that has been rounded off due to the high frequency cut-off, and with long low frequency rolling tails due to the low frequency cut-off.
The waveform is symmetric about \(t = 0\), making it acausal:  spacetime begins oscillating prior to the passage of the source of the wavefront.
Simply setting the waveform to 0 for \(t < 0\) imposes exact causality, and because of the simple symmetric shape the total energy in the wavefront can be maintained by scaling the remaining half of the waveform's amplitude by a factor of \(\sqrt{2}\), but applying this kind of sharp time domain window function also alters the spectrum undesirably by adding long polynomially decaying tails in the frequency domain that reintroduce the high-frequency energy divergence.
Instead, a half Gaussian taper is used for \(t < 0\), whose width is comparable to the time scale of the high frequency cut-off so that the taper's shape mimics the already rounded-off shape of the tip.
The waveform is not exactly causal but the rise time is very rapid and the shape of the spectrum is well preserved, and in particular the two frequency domain regularizations are preserved.
Instead of a simple factor of \(\sqrt{2}\) the amplitude normalization is adjusted numerically.

\begin{figure}
\resizebox{\linewidth}{!}{\includegraphics{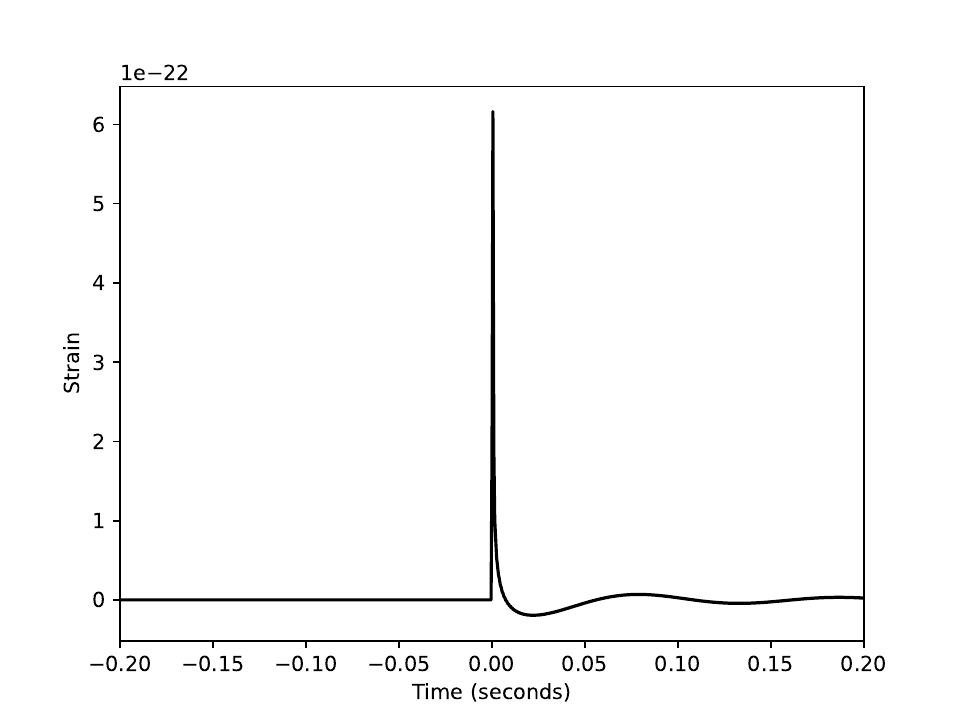}}
\caption{The Cherenkov gravitational-wave burst waveform in the time domain.}
\label{fig:waveform}
\end{figure}

An eaxmple of the final waveform model is shown in \figref{fig:waveform}.

\section{Search Method}
\label{sec:search_method}

For any given candidate event, there are two point hypotheses:  ``noise'' and ``signal + noise''.
We use the logarithm of an approximation of the likelihood ratio to rank candidate events from most signal-like to least signal-like.
\begin{equation}
        \ln \likelihood = \ln \frac{P(\text{data}|\text{signal})}{P(\text{data}|\text{no signal})},
\end{equation}
where $P(\text{data}|\text{no signal})$ is the probability of obtaining data given the null hypothesis and $P(\text{data}|\text{signal})$ is that given the hypothesis that there is a signal.
Thresholding on the likelihood ratio, \(\likelihood\), satisfies the Neyman-Pearson criterion \cite{NeymanPearson}, providing the highest detection efficiency at fixed false alarm rate.
Any other function that is monotonic in \(\likelihood\) is equivalent, and we use \(\ln \likelihood\).

Candidates are identified in the \ac{GW} detector strain time series using a matched-filter \ac{SNR} \cite[equation (A24)]{cutler1994} peak-finding algorithm borrowed from the search for \ac{GW} bursts from cosmic string cusps \cite{siemens2006}.  The filter bank consists of the single template described in Sec.~\ref{sec:template}.

For the parameters used to define the ranking statistic, we choose the magnitude of the peak of the matched-filter \ac{SNR} time series, $\rho$, and a $\chi^2$-like waveform consistency measure, which we denote by $\xi^2$.
$\xi^2$ is a weighted sum-of-square residuals obtained from a fit of the template autocorrelation function to the \ac{SNR} time series.  It was originally developed for searches for \acp{GW} from compact object mergers, it has been used in searches for \ac{GW} bursts from cosmic string cusps \cite{PhysRevLett.126.241102}, and is described in more detail in \cite{Cody2017FrameworkCBMerger}.
Explicity, the four-parameter ranking statistic has the form
\begin{equation}
        \ln \likelihood(\rho_{\Hanford},\xi^2_{\Hanford},\rho_{\Livingston},\xi^2_{\Livingston}) = \ln \frac{P(\rho_{\Hanford},\xi^2_{\Hanford},\rho_{\Livingston},\xi^2_{\Livingston}|\text{signal})}{P(\rho_{\Hanford},\xi^2_{\Hanford},\rho_{\Livingston},\xi^2_{\Livingston}|\text{noise})}.
\end{equation}

Assuming the noise processes at Hanford and Livingston to be uncorrelated, the denominator can be factored
\begin{multline}
        \label{eq:denominator}
P(\rho_{\Hanford},\xi^2_{\Hanford},\rho_{\Livingston},\xi^2_{\Livingston}|\text{noise}) = \\ P(\rho_{\Hanford},\xi^2_{\Hanford}|\text{noise}) \cdot P(\rho_{\Livingston},\xi^2_{\Livingston}|\text{noise}).
\end{multline}
Estimates of $P(\rho_{\Hanford},\xi^2_{\Hanford}|\text{noise})$ and $P(\rho_{\Livingston},\xi^2_{\Livingston}|\text{noise})$ are obtained by applying kernel density estimation to $(\rho, \xi^2)$ samples drawn from candidates.
When this technique is used for compact object searches, candidates that are seen in coincidence by several observatories are excluded from the \acp{PDF} otherwise samples that are the result of genuine signals could contaminate the \acp{PDF} and diminish the search's ability to differentiate signals from noise.
Application of that criterion requires that samples only be collected during times when at least two detectors are operating, which limits the data available for the procedure.
Since we have only one template, the rate at which \((\rho, \xi^{2})\) samples are obtained is much slower, so to avoid not collecting enough triggers for the \ac{PDF} estimates to converge, and because we expect we are in the strongly noise-dominated regime, we used all single detector triggers for the denominator and accept a small potential loss of sensitivity.

As was done in \cite{cannon2014a}, we begin by factoring the numerator as
\begin{multline}
P(\rho_{\Hanford},\xi^2_{\Hanford},\rho_{\Livingston},\xi^2_{\Livingston}|\text{signal}) = P(\rho_{\Hanford},\rho_{\Livingston}|\text{signal})\cdot\\ P(\xi_{\Hanford}^2|\rho_{\Hanford},\text{signal})\cdot P(\xi_{\Livingston}^2|\rho_{\Livingston},\text{signal}).
\end{multline}
This takes advantage of the observation that apart from their correlation with \ac{SNR}, which for signals is correlated across multiple detectors, the \(\xi^2\) are statistically independent of each other.
We approximate this expression with
\begin{multline}
        \label{eq:numerator}
P(\rho_{\Hanford},\xi^2_{\Hanford},\rho_{\Livingston},\xi^2_{\Livingston}|\text{signal}) \approx \\ P(\rho_{\Hanford}|\text{signal}) \cdot P(\rho_{\Livingston}|\text{signal}) \cdot\\ P(\xi_{\Hanford}^2|\rho_{\Hanford},\text{signal})\cdot P(\xi_{\Livingston}^2|\rho_{\Livingston},\text{signal}),
\end{multline}
\textit{i.e.}\ we ignore \ac{SNR} correlations among the detectors.
This approximation of the numerator greatly simplifies the analysis, and has no scientific consequences other than diminishing the quality of the ranking statistic by departing slightly from the true likelihood ratio.  Since we have no expectation of making a detection, the cost in software development time to do this better is not warranted, but this is an obvious avenue for improvement.

\begin{figure}
\resizebox{\linewidth}{!}{\includegraphics{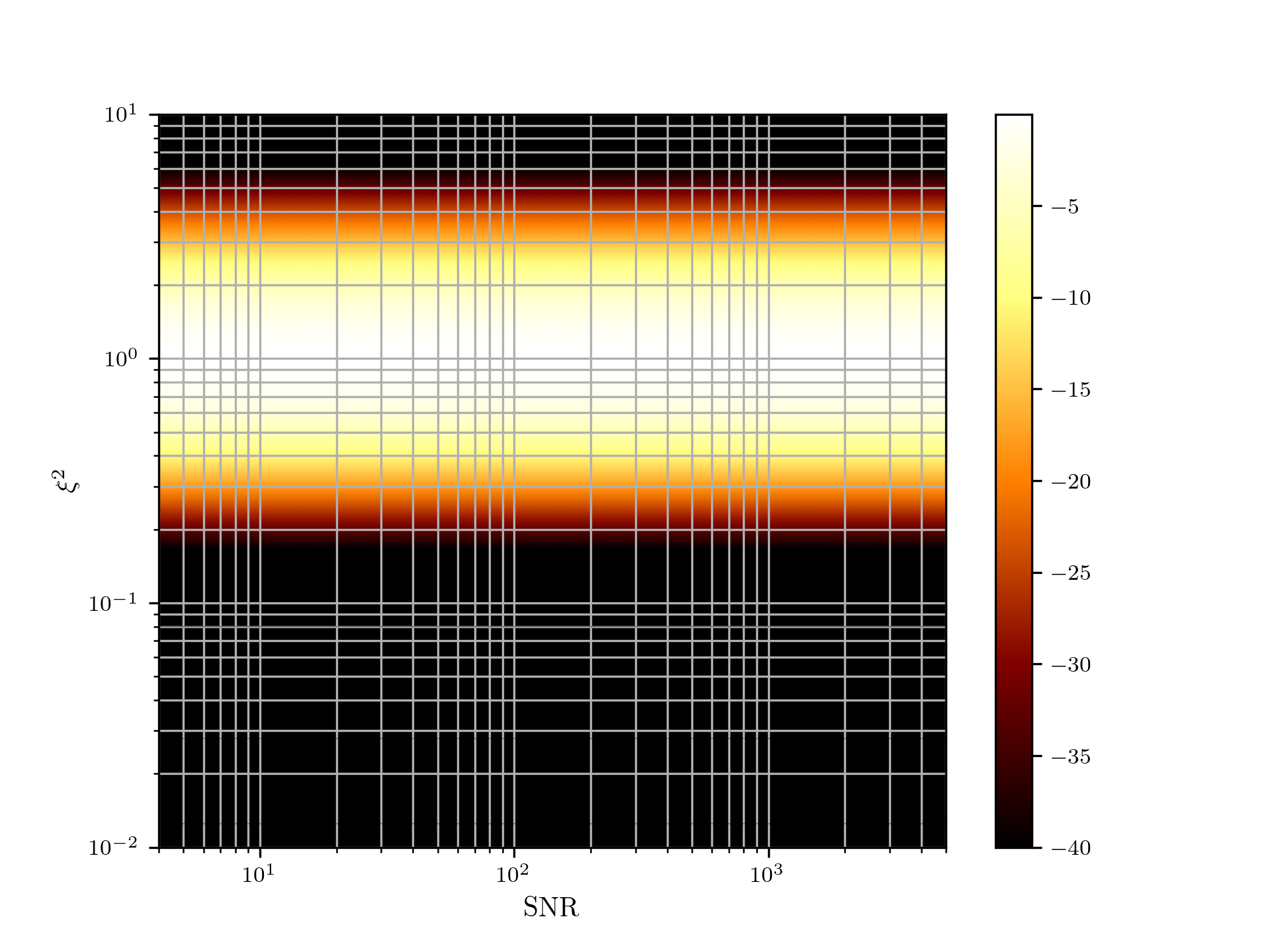}}
\caption{$\ln P(\xi^2|\rho,\text{signal})$, same for both detectors.}
\label{fig:numPDF}
\end{figure}
Based on studies of candidates obtained by adding simulated signals to detector data, $P(\xi^2|\rho,\text{signal})$ is chosen to be proportional to a Gaussian distribution in logarithmic space on $\xi^2$ axis around $\xi^2=1$. \figref{fig:numPDF} shows the \ac{PDF} plot of $\ln P(\xi^2|\rho,\text{signal})$ for Hanford detector. 

To obtain $P(\rho|\mathrm{signal})$, we first note that \ac{SNR}, $\rho$, is proportional to \ac{GW} strain, $h$.  From \eqref{eq:waveform}, $h^2$ is proportional to $r^{-1}$, therefore
\begin{subequations}
\label{eq:drdrho}
\begin{align}
r & \propto \rho^{-2}\\
\diff r & \propto \rho^{-3}\diff\rho
\end{align}
\end{subequations}
Assuming sources are uniformly distributed in volume, the number of them in a spherical shell of thickness $\diff r$ at some distance $r$ is $\propto r^{2} \diff r$.  Therefore, together with \eqref{eq:drdrho},
\begin{equation}
P(\rho|\mathrm{signal})\diff\rho \propto r^2\diff r \propto\rho^{-7}\diff\rho
\label{eq:Prho-7}
\end{equation}

\begin{figure}
\resizebox{\linewidth}{!}{\includegraphics{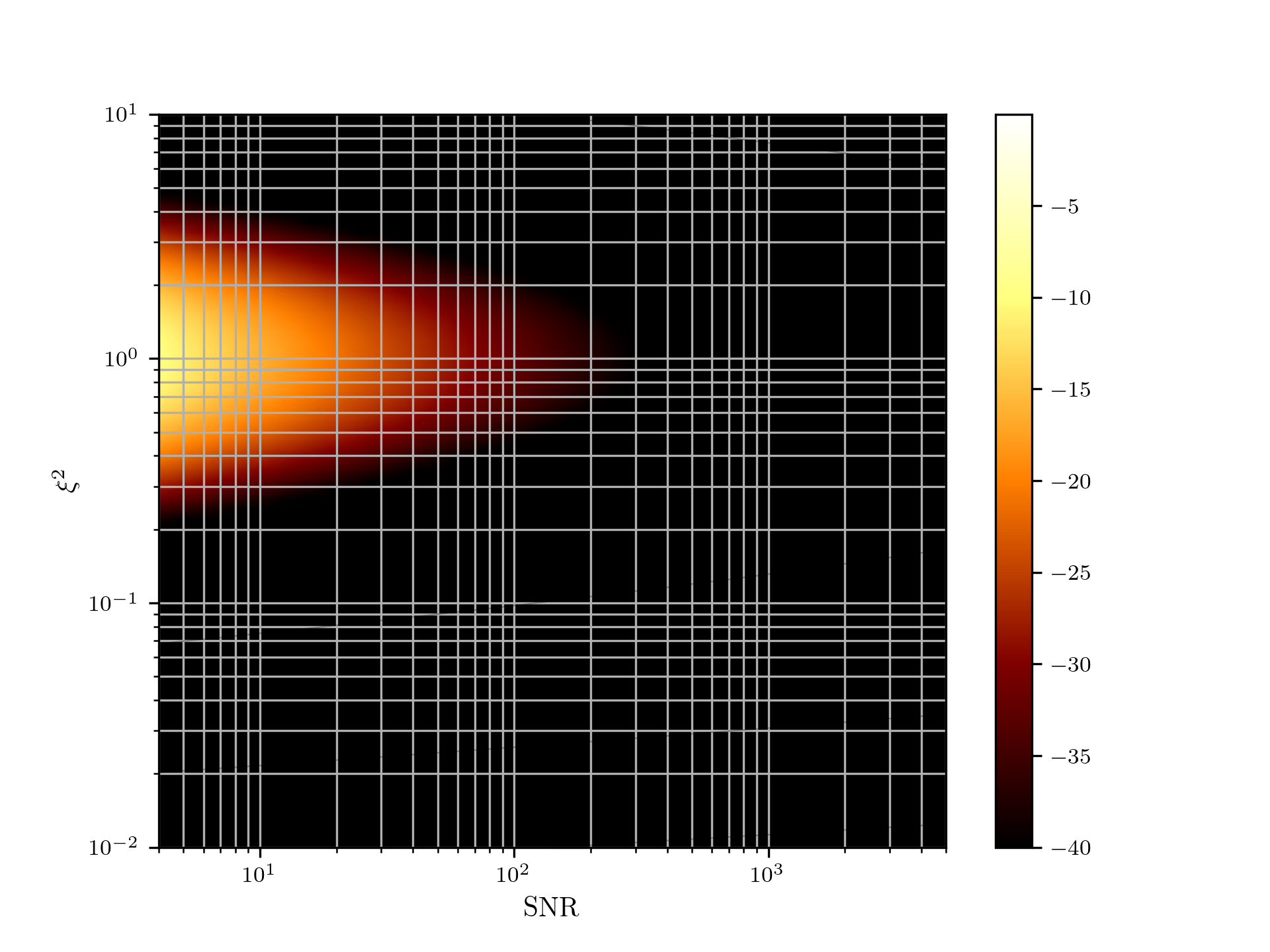}}
\caption{$\ln P(\rho|\mathrm{signal}) + \ln P(\xi^2|\rho,\mathrm{signal})$, same for both detectors.}
\label{fig:numPDFrho7}
\end{figure}
Therefore we can obtain $\ln P(\xi^2,\rho|\text{signal})$ by calculating $P(\rho|\mathrm{signal})\cdot P(\xi^2|\rho,\mathrm{signal})$ as is shown in \figref{fig:numPDFrho7}. For the normal isotropic radiation case, $P(\rho|\mathrm{signal}) \propto \rho^{-4}$, and so, perhaps counter-intuitively for the reader (it was for us), the fact that the strain amplitude falls off more slowly with distance for this waveform leads to an \ac{SNR} distribution that much more strongly favours \emph{lower} \ac{SNR} signals when compared to an isotropic radiation model, not higher \ac{SNR}.  $P(\rho|\mathrm{signal})$ is the relative frequency with which various \acp{SNR} are observed, and, assuming a uniform distribution, there are many more distant sources than nearby sources, so changing the power-law relationship between distance and \ac{SNR} to allow signals from greater distances to be seen tilts the relative frequencies of \ac{SNR} to smaller values.

One benefit of this outcome for this search is that the signal model in the ranking statistic numerator provides a natural glitch veto:  high amplitude candidates are rejected by the ranking statistic, because low amplitude signals should be so much more numerous if signals are present in the data.  This effect does not occur in searches for sources that radiate spherically, like compact object mergers, where more effort must be spent mitigating the effects of non-stationary terrestrial noise artifacts in the data.

\section{Result and Discussion}
\label{sec:result}
\begin{figure}
\resizebox{\linewidth}{!}{\includegraphics{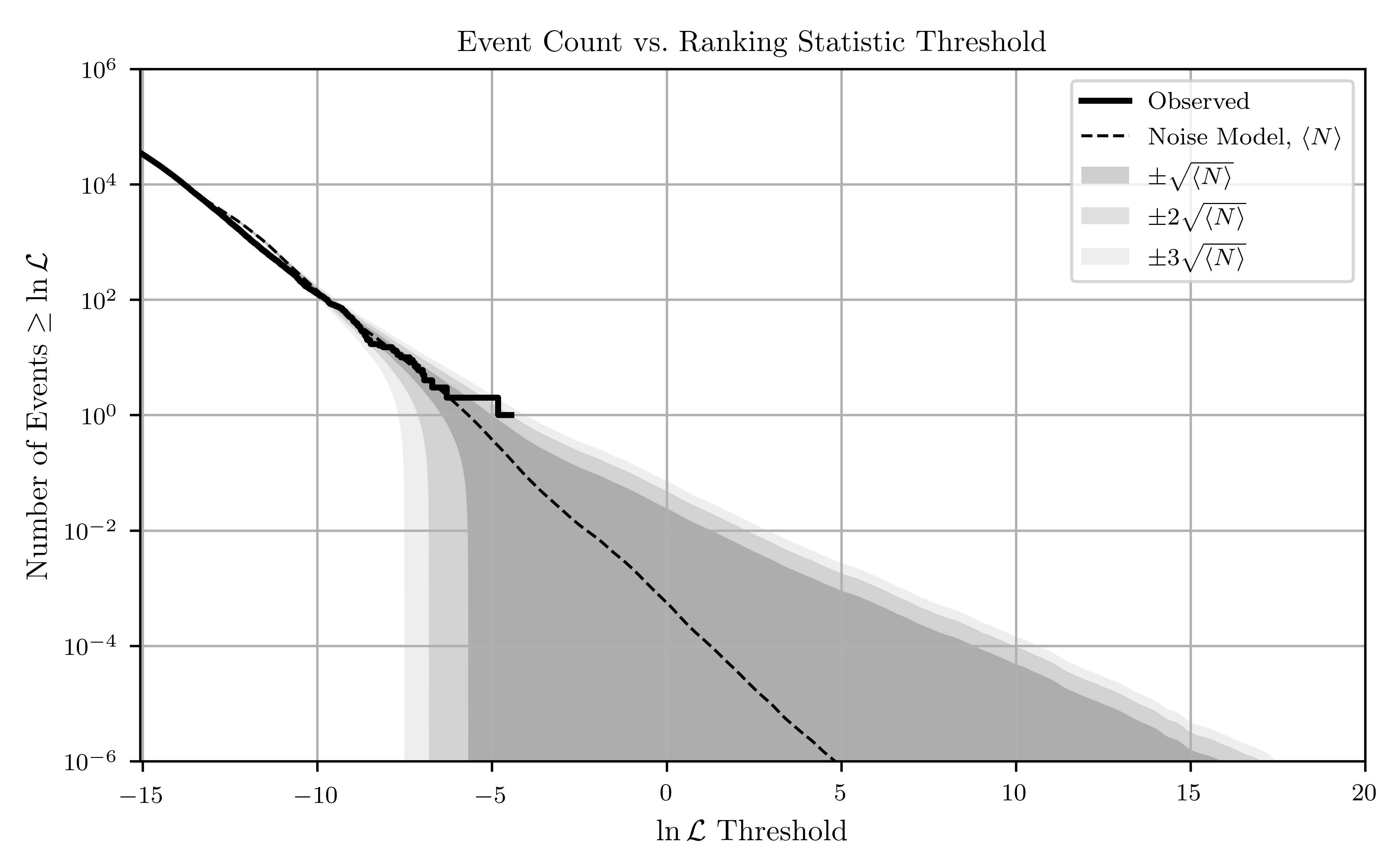}}
\caption{Summary of search result.  The observed candidate event rate as a function of ranking statistic threshold, together with the rate predicted by the noise model.  The two are consistent:  a null result.  The shaded regions show multiples of $\pm\sqrt{\left<N\right>}$ to give an indication of the scale of expected Poisson counting fluctuations.}
\label{fig:singlecollect}
\end{figure}
\figref{fig:singlecollect} is a plot obtained by running pipeline for the data from Mon Apr 01 15:00:00 GMT 2019 to Fri Mar 27 17:00:00 GMT 2020 on the third observing run. 
The data set we analyzed is from GWOSC's open data for Hanford and Livingston \cite{GWdatareference}.

\subsection{Loudest Event}

The false alarm probability for the loudest event with $\ln\likelihood=-4.44$ is 15\%.
The further investigation on the loudest two coincident events is discussed here.
\begin{figure*}
\resizebox{0.5\linewidth}{!}{\includegraphics[width=\linewidth]{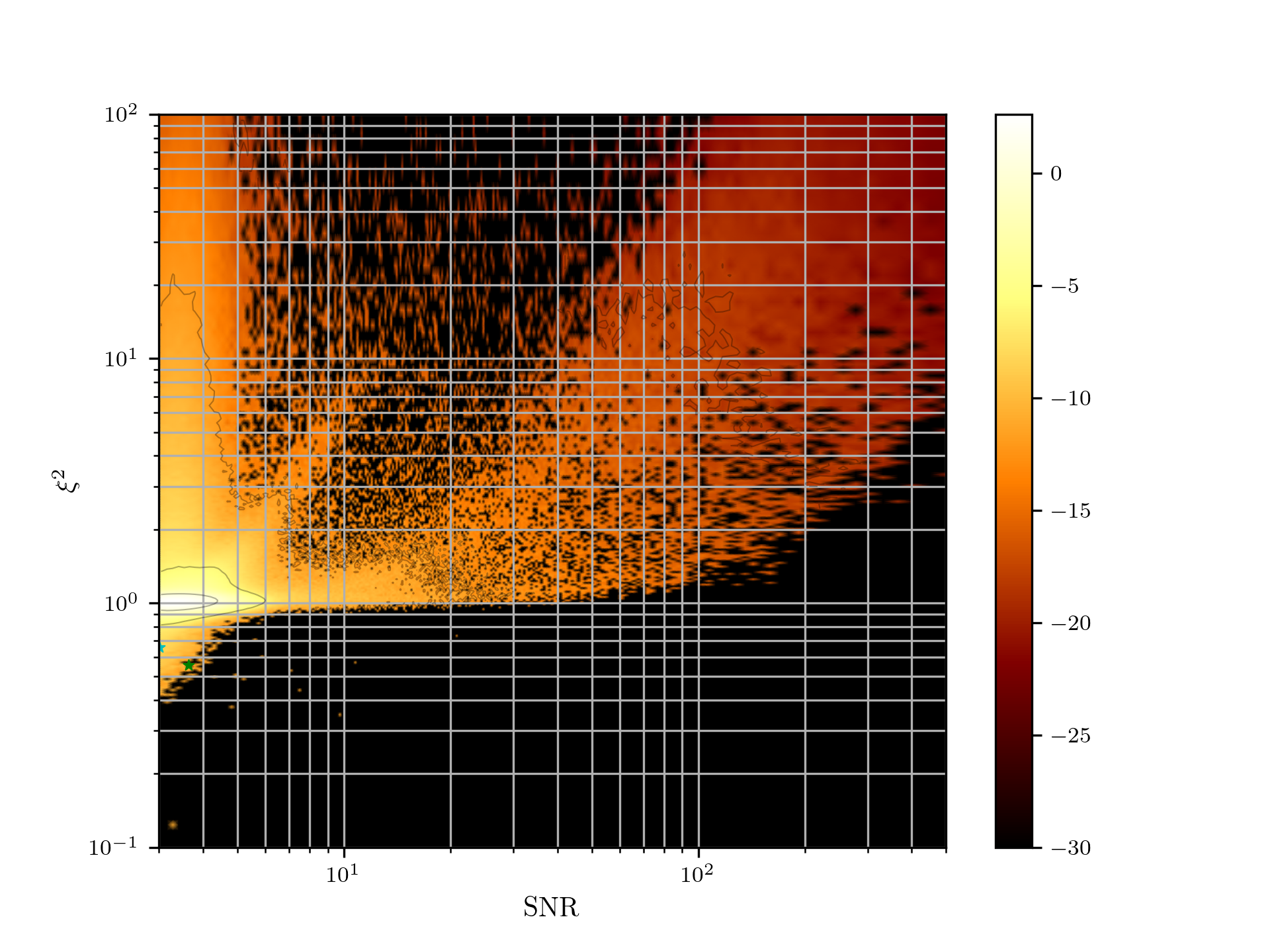}}%
\resizebox{0.5\linewidth}{!}{\includegraphics[width=\linewidth]{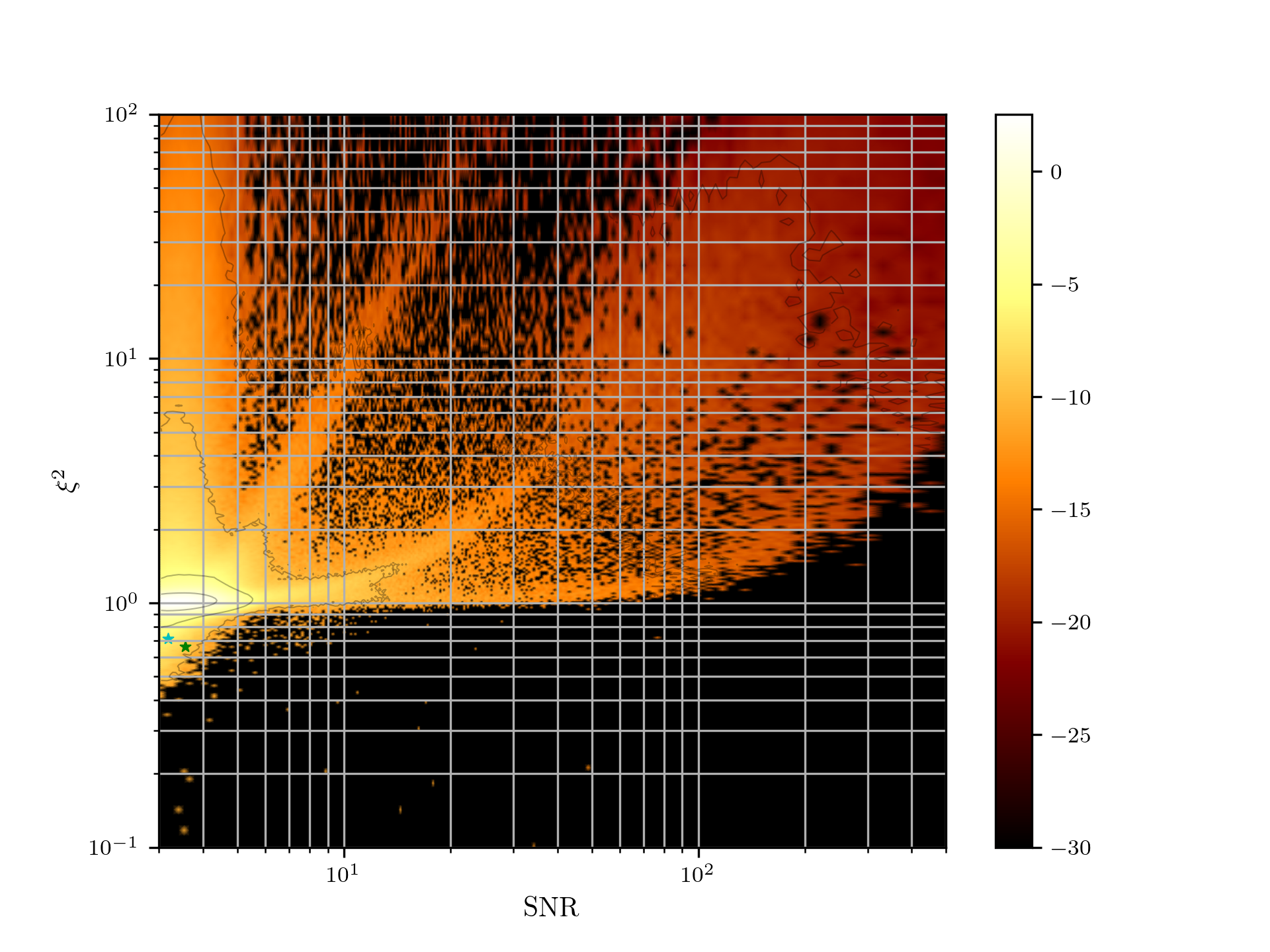}}
\caption{The denominator \ac{PDF} on \ac{SNR}-$\xi^2$ plane. The \ac{PDF} for a detector in Hanford is shown in left panel and Livingston in right panel. Two colored star point is the single events which contribute to the loudest two coincident events. The cyan star corresponds to the loudest event with $\ln\likelihood=-4.44$ and green star to the second loudest event with $\ln\likelihood=-4.82$.}
\label{fig:denPDF}
\end{figure*}
In \figref{fig:denPDF}, single events which contribute to the loudest two coincident events.
Since none of candidates indicates high \ac{SNR} and low $\xi^2$ value as shown in \figref{fig:denPDF}, the search cannot confirm the detection.
One of the plausible reason for obtaining loud event is that the event is located at the hole of the denominator \ac{PDF}.
As is mentioned in Sec.~\ref{sec:CherenkovSpectrum}, in this search, only one template was used resulting small number of events which can be used for noise(denominator) \ac{PDF}.
The application of ``kernel density estimation'' to acquire continuous \ac{PDF} on \ac{SNR}-$\xi^2$ plane did not function enough to exclude holes in the \ac{PDF} and make some coincident events which is unlikely the signal loud.
The discrepancy between noise model and zero-lag coincidents around $\ln\likelihood=-12$ on \figref{fig:singlecollect} can also be explained by same reason.

\subsection{Detection Efficiency}
To interpret the nondetection result, we can apply the loudest-event rate upper-limit method introduced in \cite{Truerate}.
The highest-ranked event from the search defines a ranking statistic threshold, $\likelihood{\star}$, and we must obtain the probability of recovering a Cherenkov burst signal above this threshold,
\begin{equation}
\label{eq:efficiency}
\epsilon ( \diff E / \diff A )
   = P \left( \likelihood \geq \likelihood{\star} | \text{signal}, \frac{\diff E}{\diff A} \right).
\end{equation}
The flux density is a convenient parameterization of the efficiency because from \eqref{eq:waveform} we see that when written in terms of flux density (and the high frequency cut-off) the waveform model does not depend on any other parameters of the problem like the distance to the source or its speed.

\begin{figure}
\resizebox{\linewidth}{!}{\includegraphics{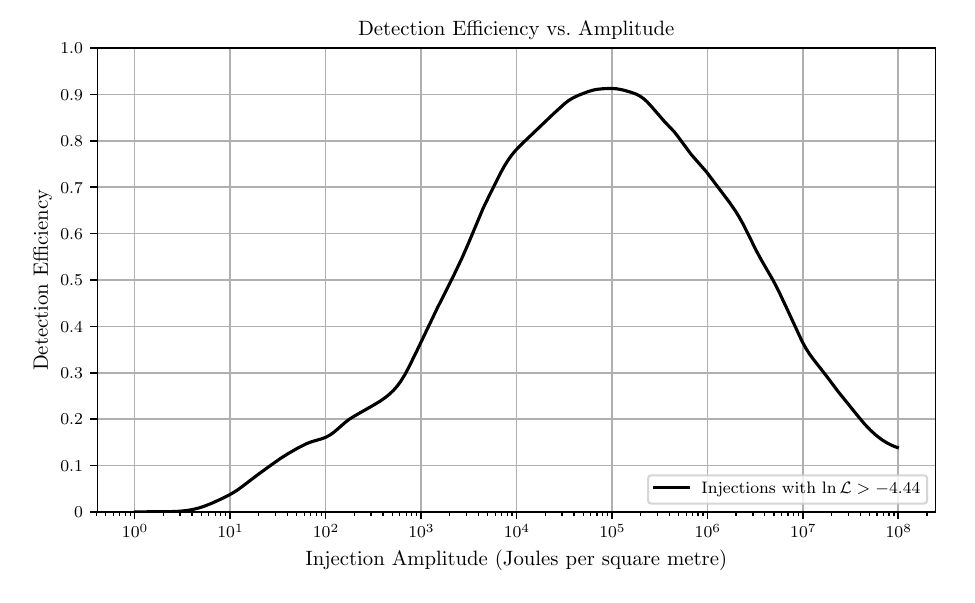}}
\caption{The detection efficiency as a function of flux density in the range $\unit{10^0}{\joulepersquaremetre}$ to $\unit{10^8}{\joulepersquaremetre}$.}
\label{fig:efficiency}
\end{figure}

We measure the detection efficiency above threshold using a Monte Carlo approach.
Simulated signals are generated with a variety of flux densities, and injected into data as if arriving at Earth from a variety of directions.  The data containing the injections are analyzed, and if a signal candidate above $\likelihood{\star}$ is identified within some time window of a given injection then that injection is ``detected'', otherwise it is ``missed''.  By binning the injections by flux density and counting the fraction within the bin that are detected one can estimate $\epsilon (\diff E / \diff A)$.
The result is shown in \figref{fig:efficiency}.

Note that the detection efficiency becomes small for high amplitude signals.  Searches for sources that radiate spherically, for example compact object mergers, asymptote to detection efficiencies near 1 for high amplitude signals.  This difference arises from the way $P(\rho|\mathrm{signal})\diff\rho$ more strongly suppresses large \ac{SNR} signals for this signal model.
Although it seems undesirable for high amplitude signals to be undetectable by the pipeline, we will see in Sec.~\ref{subsec:constraints} below that the effect only becomes significant for uninterestingly loud signals.

\subsection{Constraints}
\label{subsec:constraints}
From the measured detection efficiency and the observed absence of signals above $\likelihood{\star}$, using \cite[equation (7)]{Truerate} we can say that with 90$\%$ confidence the true rate of such signals is less than
\begin{align}
R_{90\%} ( \diff E / \diff A )
   \leq \frac{3.890}{T \epsilon( \diff E / \diff A ) }
\end{align}
where $T$ is the total observation time, and $\epsilon( \diff E / \diff A )$ is the detection efficiency measured in \eqref{eq:efficiency}.  \eqref{eq:Efunc} can be used to compute $\diff E / \diff A$ from the impact parameter, $r$, the dimensionless speed of the source, $\beta$, and the output power of the source, $P$, thereby providing the rate upper bound in terms of these three quantities.

\subsubsection{Example:  NCC-1701-D Enterprise}

In the television series Star Trek:  The Next Generation, the characters explore the galaxy aboard the space craft NCC-1701-D Enterprise.  This fictional vehicle is capable of faster-than-light travel.
The series makes varying claims about the power output of the vehicle's engines, but possibly the most specific claim is season 6, episode 6, ``True Q'', wherein the character Commander Data, in response to another character's marvelling at the immense power output of the engines, states ``Imagination is not necessary.
The scale is readily quantifiable.  We are currently generating 12.75 billion gigawatts per  [cut off by sound of alarm]''.

Per what?

\begin{figure*}
\resizebox{\linewidth}{!}{\includegraphics{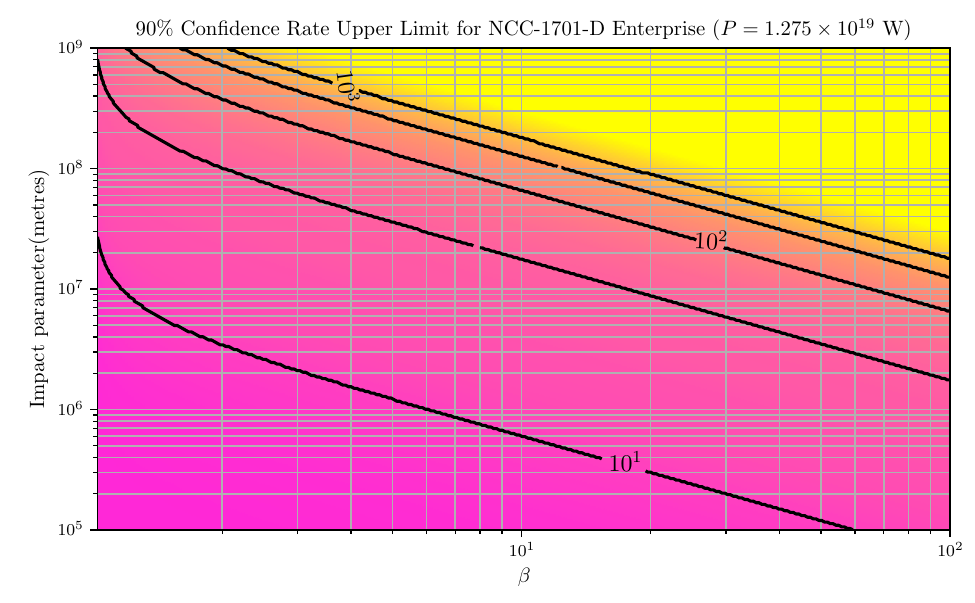}}
\caption{The 90\% confidence upper limit on event count per experiment.  The vertical axis represents impact parameter which corresponds to the distance between the observatory and trajectory of the space craft as is shown in \figref{fig:Cherenkov1r}. The horizontal axis represents $\beta$ which corresponds to the rate of the velocity of the source to the speed of light in vacuum.}
\label{fig:constraints}
\end{figure*}

Conservatively assuming the line was to continue by naming something the ship has only one of, we fix the output power of the source to be $P = \unit{12.75\times10^{18}}{\watt}$.  The vehicle is said to be powered by a matter-antimatter annihiliation reaction, but even in mass units this power output corresponds to over $\unit{140}{\kilo\gram\per\second}$, which, being a deep space exploration vessel, suggests an enormous fuel capacity.  In any case, adopting this number, we can use the null result obtained above to constrain the number of such vehicles that passed near Earth during the observation period as a function of the speed and impact parameter.
The result is shown in \figref{fig:constraints}.

For example, if we consider the vehicles passed Earth at about twice the speed of light, at $\beta \approx 2$, and at a distance of about 1/4 of the distance to the moon, at $r = \unit{10^{8}}{\metre}$, we can read off a rate upper limit of about 30 from \figref{fig:constraints}.  This tells us that had 30 or more such fly-bys occured during the period April, 2019, through March, 2020, with 90\% probability we would have detected at least one of them --- and we did not, so that rate of visits by that sort of vehicle can be excluded for that observation period.

\figref{fig:constraints} shows a trend towards more relaxed rate constraints for higher speed vehicles.  This is a consequence of assuming a fixed power output:  a faster moving source therefore spreads each unit of energy over a larger area of wavefront, lowering the flux denxity of the wave and diminshing its detectability.

We also see a trend towards tighter rate constraints for smaller impact parameters, which is obviously due to higher detection efficiency for higher amplitude, nearby, sources.  However, recalling \figref{fig:efficiency}, the detection efficiency becomes small again for very high amplitude signals, so we expect to see a weakening of the rate constraint for small impact parameters.  Since that isn't seen here, evidently that effect is not yet relevant even for impact parameters less than the radius of the Earth, and so that the loss of detection efficiency at high amplitudes has no practical consequences for a search for these signals.

\acknowledgments

This work has been supported by Japan Society for the Promotion of Science (JSPS) Grants-in-Aid for Scientific Research (KAKENHI) grant number 18H03698.

The authors are grateful for computational resources provided by the LIGO Laboratory and supported by National Science Foundation Grants PHY-0757058 and PHY-0823459.

This material is based upon work supported by NSF's LIGO Laboratory which is a major facility fully funded by the National Science Foundation.

This search has made use of gstlal software \cite{gstlal} and lalsuite software \cite{lalsuite}.

This research has made use of data or software obtained from the Gravitational Wave Open Science Center (gw-openscience.org), a service of LIGO Laboratory, the LIGO Scientific Collaboration, the Virgo Collaboration, and KAGRA. LIGO Laboratory and Advanced LIGO are funded by the United States National Science Foundation (NSF) as well as the Science and Technology Facilities Council (STFC) of the United Kingdom, the Max-Planck-Society (MPS), and the State of Niedersachsen/Germany for support of the construction of Advanced LIGO and construction and operation of the GEO600 detector. Additional support for Advanced LIGO was provided by the Australian Research Council. Virgo is funded, through the European Gravitational Observatory (EGO), by the French Centre National de Recherche Scientifique (CNRS), the Italian Istituto Nazionale di Fisica Nucleare (INFN) and the Dutch Nikhef, with contributions by institutions from Belgium, Germany, Greece, Hungary, Ireland, Japan, Monaco, Poland, Portugal, Spain. The construction and operation of KAGRA are funded by Ministry of Education, Culture, Sports, Science and Technology (MEXT), and Japan Society for the Promotion of Science (JSPS), National Research Foundation (NRF) and Ministry of Science and ICT (MSIT) in Korea, Academia Sinica (AS) and the Ministry of Science and Technology (MoST) in Taiwan.

\bibliography{references}
\end{document}